\documentclass[preprint,12pt]{elsarticle}

\usepackage{amsmath}
\usepackage{graphicx}
\usepackage{bm}
\usepackage{slashed}
\usepackage{amssymb}

\journal{Physics Letters B}

\begin{document}

\begin{frontmatter}



\title{Proposal for measuring new transverse momentum dependent parton distributions $g_{1T}$ and $h_{1L}^\perp$ through semi-inclusive deep inelastic scattering}


\author[PKU]{Jiacai Zhu}
\author[PKU,CHEP]{Bo-Qiang Ma\corref{cor}}
\ead{mabq@pku.edu.cn}

\address[PKU]{School of
Physics and State Key Laboratory of Nuclear Physics and Technology,
Peking University, Beijing 100871, China}

\address[CHEP]{Center for High Energy
Physics, Peking University, Beijing 100871,
China}

\cortext[cor]{Corresponding author.}

\begin{abstract}
We calculate $g_{1T}$ and $h_{1L}^\perp$, two of the eight leading
twist transverse momentum dependent parton distributions (TMDs), in
the light-cone quark-diquark model. The new TMDs can be measured
through semi-inclusive deep inelastic scattering (SIDIS). We present
predictions of the single and double spin asymmetries related to
$g_{1T}$ and $h_{1L}^\perp$ in SIDIS at HERMES, COMPASS, and JLab
kinematics respectively.
\end{abstract}


\end{frontmatter}



\section{Introduction}
Transverse momentum dependent parton distributions
(TMDs)~\cite{Barone2002,Barone2010}, as a generalization of parton
distribution functions (PDFs) from one dimension to three dimensions
in momentum space, provide rich information on nucleon structure. At
leading twist, there are eight TMDs contained in the quark-quark
correlation matrix~\cite{Mulders1996,Boer1998}:
\begin{equation}
\begin{split}
\Phi(x, \bm{k}_T) = & \frac{1}{2} \bigg\{ f_1\slashed{n}_+ - f_{1T}^\perp \frac{\epsilon_T^{ij} k_{Ti} S_{Tj}}{M_N}\slashed{n}_+ + \Big(S_L g_{1L} + \frac{\bm{k}_T \cdot \bm{S}_T}{M_N} g_{1T}\Big) \gamma_5\slashed{n}_+ \\
& \quad + h_{1T}\frac{[\slashed{S}_T,\slashed{n}_+]\gamma_5}{2} + \Big(S_L h_{1L}^\perp + \frac{\bm{k}_T \cdot \bm{S}_T}{M_N} h_{1T}^\perp\Big) \frac{[\slashed{k}_T,\slashed{n}_+]\gamma_5}{2M_N} \\
& \quad + i h_1^\perp \frac{[\slashed{k}_T,\slashed{n}_+]}{2M_N} \bigg\}.
\end{split}
\end{equation}
Using the notation $\Phi^{[\Gamma]} \equiv \mathrm{Tr}(\Phi\Gamma)/2$, one gets
\begin{align}
\Phi^{[\gamma^+]} =& f_1(x, k_T^2) - S_{Tj} \frac{\epsilon_T^{ij} k_{Ti}}{M_N}f_{1T}^\perp(x, k_T^2),\\
\Phi^{[\gamma^+\gamma_5]} =& S_L g_{1L}(x, k_T^2) + \frac{\bm{k}_T \cdot \bm{S}_T}{M_N} g_{1T}(x, k_T^2),\\
\Phi^{[i\sigma^{i+}\gamma_5]} =& S^i_T h_1(x, k_T^2) + S_L \frac{k^i_T}{M_N} h_{1L}^\perp(x, k_T^2)
- S_{Tj}\frac{2k^i_T k^j_T + k^2_T g_T^{ij}}{2M_N^2}h_{1T}^\perp(x, k_T^2) \nonumber\\
&- \frac{\epsilon_T^{ij} k_{Tj}}{M_N} h_1^\perp(x, k_T^2), \quad i=1, 2,
\end{align}
where $h_1(x, k_T^2) = h_{1T}(x, k_T^2) + (k_T^2 / 2M_N^2) h_{1T}^\perp(x, k_T^2)$.
If we integrate over transverse momenta $\bm{k}_T$, these three correlation
functions $\Phi^{[\gamma^+]}$, $\Phi^{[\gamma^+\gamma_5]}$, and
$\Phi^{[i\sigma^{i+}\gamma_5]}$ reduce to the unpolarized, helicity,
and transversity distributions respectively, while other five TMDs vanish.

Among the eight TMDs, $g_{1T}$ and $h_{1L}^\perp$ are probably the
least considered ones. Besides, they reflect new information on the
quark spin and orbital correlation of the nucleon. $g_{1T}$
describes the probability of finding a longitudinally polarized
quark inside a transversely polarized nucleon, so we could call it
transversal helicity (or shortly, trans-helicity). In the same
manner, $h_{1L}^\perp$ could be called longitudinal transversity (or
shortly, longi-transversity or heli-transversity) for that it
represents the probability of finding a transversely polarized quark
inside a longitudinally polarized nucleon. The unique feature of
$g_{1T}$ and $h_{1L}^\perp$ is that for other six TMDs, there are
corresponding generalized parton distributions (GPDs) in the
light-cone quark models, but for $g_{1T}$ and $h_{1L}^\perp$, the
corresponding GPDs vanish because of time
invariance~\cite{Diehl2005}.

$h_{1L}^\perp$ is chiral-odd, so it can be probed via SIDIS when
combined with the chiral-odd Collins fragmentation function. For $g_{1T}$,
it is chiral-even, and can be measured through SIDIS when combined with
the unpolarized fragmentation function.

\section{$g_{1T}$ and $h_{1L}^\perp$ in the light-cone quark-diquark model}
In the light-cone quark-diquark model~\cite{Ma1996,Ma2000}, if any
one of the quarks in the proton is struck, the other parts of the
proton can be effectively treated as a spectator with its quantum
numbers being those of a diquark with spin 0 or 1 (scalar and vector
diquarks). Moreover, the Melosh-Wigner rotation~\cite{Wigner1939,
Melosh1974}, which plays an important role to explain the proton spin
puzzle~\cite{Ma1991,Ma1993} due to the relativistic effect of quark transversal motions, is taken into account in the model. In
Ref.~\cite{She2009}, the pretzelosity distribution $h_{1T}^\perp$
was calculated in the light-cone quark-diquark model. Following the
same method 
with the proper selections of the proton polarization
directions, we get the expressions of $g_{1T}$ and $h_{1L}^\perp$ in
the light-cone quark-diquark model:
\begin{align}
g_{1T}^{(uv)} =& -\frac{1}{16\pi^3} \times (\frac{1}{9} \sin^2\theta_0 \varphi_V^2 W_V^g - \cos^2\theta_0 \varphi_S^2 W_S^g),\nonumber\\
g_{1T}^{(dv)} =& -\frac{1}{8\pi^3} \times \frac{1}{9} \sin^2\theta_0 \varphi_V^2 W_V^g, \label{eq:g1}
\end{align}
and
\begin{align}
h_{1L}^{\perp(uv)} =& -\frac{1}{16\pi^3} \times (\frac{1}{9} \sin^2\theta_0 \varphi_V^2 W_V^h - \cos^2\theta_0 \varphi_S^2 W_S^h),\nonumber\\
h_{1L}^{\perp(dv)} =& -\frac{1}{8\pi^3} \times \frac{1}{9} \sin^2\theta_0 \varphi_V^2 W_V^h, \label{eq:h1}
\end{align}
with the Melosh-Wigner rotation factors ($D = V, S$):
\begin{align}
W_D^g(x, k_T^2) =& \frac{2M_N(x\mathcal{M}_D + m_q)}{(x\mathcal{M}_D + m_q)^2 + k_T^2}, \label{eq:mw_g}\\
W_D^h(x, k_T^2) =& -\frac{2M_N(x\mathcal{M}_D + m_q)}{(x\mathcal{M}_D + m_q)^2 + k_T^2}, \label{eq:mw_h}
\end{align}
where
\begin{equation}
\mathcal{M}_D = \sqrt{\frac{m_q^2 + k_T^2}{x} + \frac{m_D^2 + k_T^2}{1- x}}.
\end{equation}
$\varphi_D (D = V, S)$ is the wave function in the momentum space for
the quark-diquark, and for which we can use the Brodsky-Huang-Lepage
(BHL) prescription~\cite{Brodsky1982,Huang1994}:
\begin{equation}
\varphi_D(x, k_T^2)=A_D\exp\Big\{-\frac{1}{8\alpha_D^2}\big[\frac{m_q^2 + k_T^2}{x}+\frac{m_D^2 + k_T^2}{1-x}\big]\Big\}.
\end{equation}
The parameters $\alpha_D = 0.33 ~ \mathrm{GeV}$, the quark mass $m_q = 0.33 ~ \mathrm{GeV}$, the diquark mass
$m_S = 0.60 ~ \mathrm{GeV}$, $m_V = 0.80 ~ \mathrm{GeV}$, and $\theta_0=\pi/4$ are adopted for numerical calculation.
$\theta_0$ is the mixing angle that breaks the SU(6) symmetry when $\theta_0 \neq \pi/4$.

The unpolarized distributions, in the light-cone quark-diquark
model, can be found in Refs.~\cite{Ma1996,Ma2000}:
\begin{align}
f_1^{(uv)}(x, k_T^2) &= \frac{1}{16\pi^3}   (\frac{1}{3} \sin^2\theta_0 \varphi_V^2 + \cos^2\theta_0 \varphi_S^2),\nonumber\\
f_1^{(dv)}(x, k_T^2) &= \frac{1}{8\pi^3}  \frac{1}{3} \sin^2\theta_0 \varphi_V^2. \label{eq:unpol}
\end{align}
Using Eqs.~(\ref{eq:g1}), (\ref{eq:h1}), and (\ref{eq:unpol}), one
can express $g_{1T}$ and $h_{1L}^\perp$ with the unpolarized distributions:
\begin{align}
g_{1T}^{(uv)}(x, k_T^2) =& \big[f_1^{(uv)}(x, k_T^2) - \frac{1}{2} f_1^{(dv)}(x, k_T^2)\big] W_S^g(x, k_T^2) -\frac{1}{6} f_1^{(dv)}(x, k_T^2) W_V^g(x, k_T^2),\nonumber\\
g_{1T}^{(dv)}(x, k_T^2) =& -\frac{1}{3} f_1^{(dv)}(x, k_T^2) W_V^g(x, k_T^2),
\label{eq:g2}
\end{align}
and
\begin{align}
h_{1L}^{\perp(uv)}(x, k_T^2) =& \big[f_1^{(uv)}(x, k_T^2) - \frac{1}{2} f_1^{(dv)}(x, k_T^2)\big] W_S^h(x, k_T^2) -\frac{1}{6} f_1^{(dv)}(x, k_T^2) W_V^h(x, k_T^2),\nonumber\\
h_{1L}^{\perp(uv)}(x, k_T^2) =& -\frac{1}{3} f_1^{(dv)}(x, k_T^2) W_V^h(x, k_T^2).
\label{eq:h2}
\end{align}

From Eqs.~(\ref{eq:mw_g}) and (\ref{eq:mw_h}), one gets the following
relation:
\begin{equation}
g_{1T}^{(qv)}(x, k_T^2) = -h_{1L}^{\perp(qv)}(x, k_T^2),
\end{equation}
which is supported in other
models~\cite{Jakob1997,Pasquini2008,Efremov2009,Avakian2010}. With
expressions of $h_1$ in Refs.~\cite{Schmidt1997,Ma1998} and
$h_{1T}^\perp$ in Ref.~\cite{She2009}, we get
\begin{equation}
\frac{1}{2}\big[h_{1L}^{\perp(qv)}(x, k_T^2)\big]^2 \geqslant -h_1^{(qv)}(x, k_T^2)h_{1T}^{\perp(qv)}(x, k_T^2),
\end{equation}
where the two sides are equal only if $q = d$. This is different
from what has been obtained in other
models~\cite{Efremov2009,Avakian2010}, where the two sides are equal
for both $q = u$ and $q = d$.

\section{$g_{1T}$ and $h_{1L}^\perp$ related asymmetries in semi-inclusive deep inelastic scattering}
$g_{1T}$ and $h_{1L}^\perp$ can be measured via the double and
single spin asymmetries in the SIDIS process
respectively~\cite{Kotzinian1995,Bacchetta2007}. There have been
also other proposals to measure $g_{1T}$ through other
process~\cite{Lu2007} and $h_{1L}^\perp$ through
SIDIS~\cite{Ma2001,Ma2002}. 
The cross section of the SIDIS process reads:
\begin{align}
\frac{d\sigma}{dx \, dy \, d\psi \, dz \, d\phi_h \, d P_{h\perp}^2} =& \frac{\alpha^2}{x \, y \, Q^2}\frac{y^2}{2 \, (1-\varepsilon)} \left( 1+\frac{\gamma^2}{2 \, x} \right) \Big\{ F_{UU, T} + S_L \varepsilon \sin(2\phi_h) F_{UL}^{\sin2\phi_h}\nonumber\\
& + S_T \lambda_e  \sqrt{1 - \varepsilon^2} \cos(\phi_h - \phi_S) F_{LT}^{\cos(\phi_h - \phi_S)} + \ldots \Big\},
\end{align}
and other terms will not contribute in our analysis below. Using notations
$\bm{\hat{h}} \equiv \bm{P}_{h\perp}/P_{h\perp}$ and
\begin{equation}
\begin{split}
\mathcal{F}\left[w(\bm{k}_T,\bm{p}_T)fD\right] =& x \sum_q e_q^2 \int d \bm{k}_T ~ d \bm{p}_T ~ \delta^{(2)}(\bm{k}_T - \bm{p}_T - \bm{P}_{h\perp} / z)\\
 & \times w(\bm{k}_T,\bm{p}_T) [ f^q(x, k_T^2) D^q(z, p_T^2) + f^{\bar{q}}(x, k_T^2) D^{\bar{q}}(z, p_T^2)],
\end{split}
\end{equation}
one has
\begin{align}
F_{UU,T} &= \mathcal{F} \big[f_1D_1\big],\\
F_{LT}^{\cos(\phi_h-\phi_S)} &= \mathcal{F}\bigg[\frac{\hat{\bm{h}}\cdot\bm{k}_T}{M_N} g_{1T} D_1\bigg],\\
F_{UL}^{\sin 2\phi_h} &= \mathcal{F} \bigg[-\frac{2(\hat{\bm{h}}\cdot\bm{p}_T)(\hat{\bm{h}}\cdot\bm{k}_T)-\bm{p}_T\cdot\bm{k}_T}{M_NM_h} h_{1L}^\perp H_1^\perp \bigg].
\end{align}
We define the asymmetries related to $g_{1T}$ and $h_{1L}^\perp$ as
\begin{align}
A_{LT}^{\cos(\phi_h-\phi_S)} &= \frac{F_{LT}^{\cos(\phi_h-\phi_S)}}{F_{UU,T}},\\
A_{UL}^{\sin 2\phi_h} &= \frac{F_{UL}^{\sin 2\phi_h}}{F_{UU,T}}.
\end{align}

\begin{figure}
\centering
\includegraphics[width=0.80\textwidth]{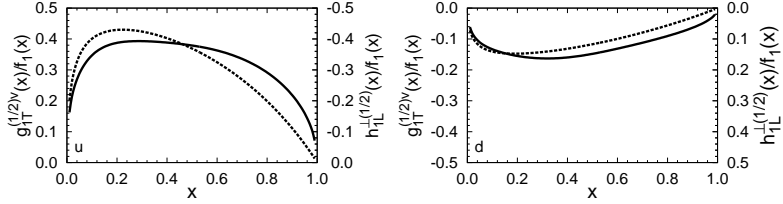}
\caption{\label{fig:ratio}The ratios $g_{1T}^{(1/2)v}(x) / f_1(x)$ and $h_{1L}^{\perp(1/2)v}(x) / f_1(x)$. Dashed curves correspond to approach 1, while solid curves correspond to approach 2. Left for $u$ quark, and right for $d$ quark.}
\end{figure}

We present numerical calculations in two different approaches: for
approach 1, we use Eqs.~(\ref{eq:g1}) and Eqs.~(\ref{eq:h1}) directly
to calculate; while for approach 2, we adopt the CTEQ6L
parametrization~\cite{Pumplin2002} for the unpolarized distributions, 
assume a Gaussian form factor of transverse momentum as suggested
in Ref.~\cite{Anselmino2009}:
\begin{equation}
f_1(x, k_T^2) = f_1(x) \frac{\exp(-k_T^2 / k_{av}^2)}{\pi k_{av}^2}
\end{equation}
with $k_{av}^2 = 0.25~\mathrm{GeV}^2$, and then use Eqs.~(\ref{eq:g2})
and Eqs.~(\ref{eq:h2}) to calculate. The ratios $g_{1T}^{(1/2)v}(x) / f_1(x)$
and $h_{1L}^{\perp(1/2)v}(x) / f_1(x)$ are shown in Fig.~\ref{fig:ratio}
with $Q^2 = 3.0~\mathrm{GeV}^2$, where
the notation
\begin{equation}
j^{(1/2)}(x) = \int d \bm{k}_T \bigg(\frac{k_T^2}{2M_N^2}\bigg)^{1/2} j(x, k_T^2)
\end{equation}
for TMD $j(x, k_T^2)$ is used. For the unpolarized and Collins
fragmentation functions, we adopt the forms suggested in the same
paper~\cite{Anselmino2009}, and the parametrization of $D_1(z)$
can be found in Ref.~\cite{Florian2007}. When we calculate the
asymmetries, we only sum over the valence quark distributions in
approach 1, while in approach 2, the unpolarized quark and antiquark
distributions are considered. As we mentioned in Ref.~\cite{She2009},
approach 2 involves the CTEQ6L parametrization which has been well
verified and constrained by many experiments, and can give more
reasonable predictions for future experiments.

\begin{figure}
\centering
\includegraphics[width=0.80\textwidth]{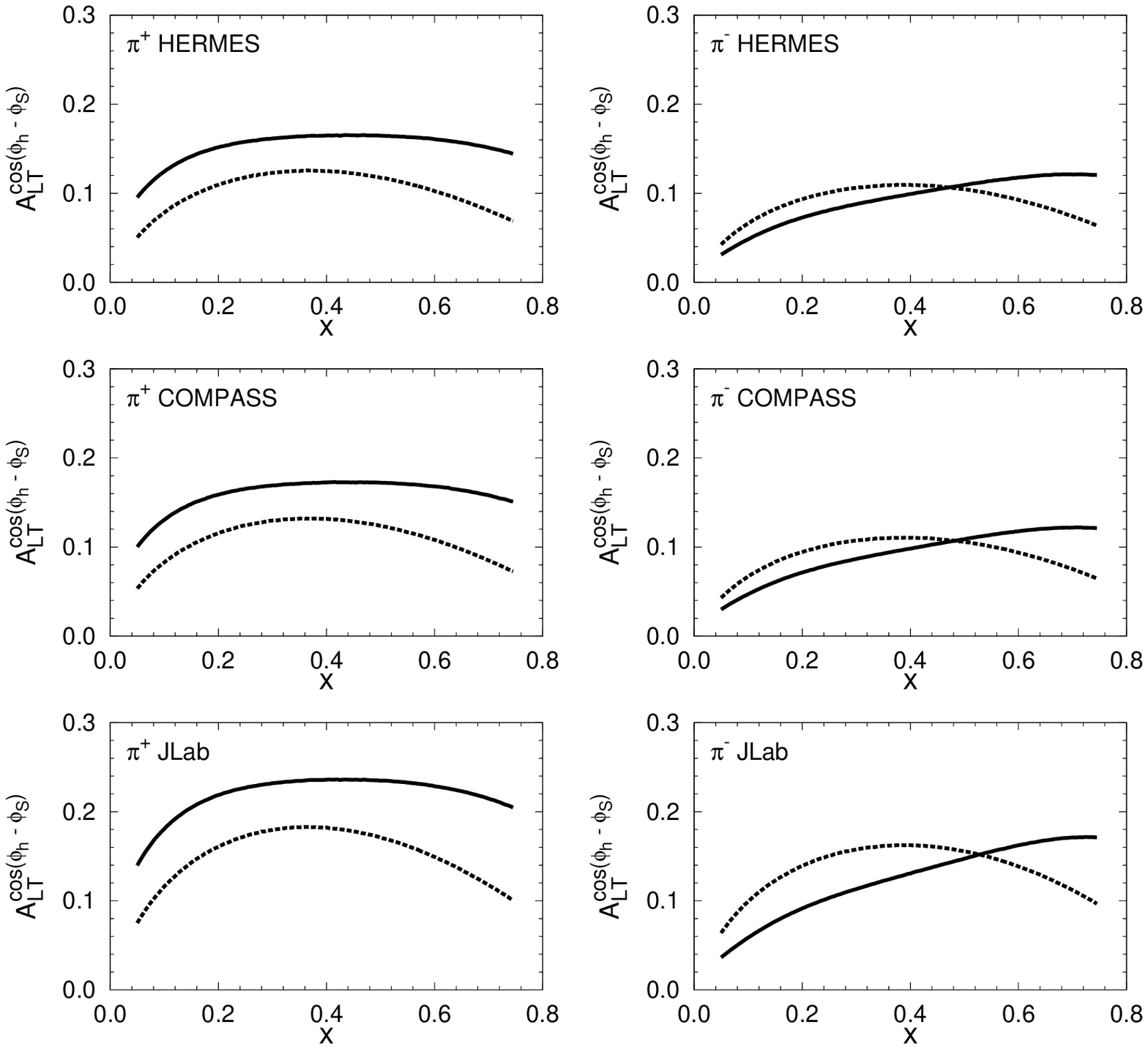}
\caption{\label{fig:g_p}The double spin asymmetry $A_{LT}^{\cos(\phi_h-\phi_S)}$ as a function of $x$ at different kinematics with $Q^2 = 3.0 ~ \mathrm{GeV}^2$ for the proton target. Dashed curves correspond to approach 1, while solid curves correspond to approach 2.}
\end{figure}

\begin{figure}
\centering
\includegraphics[width=0.80\textwidth]{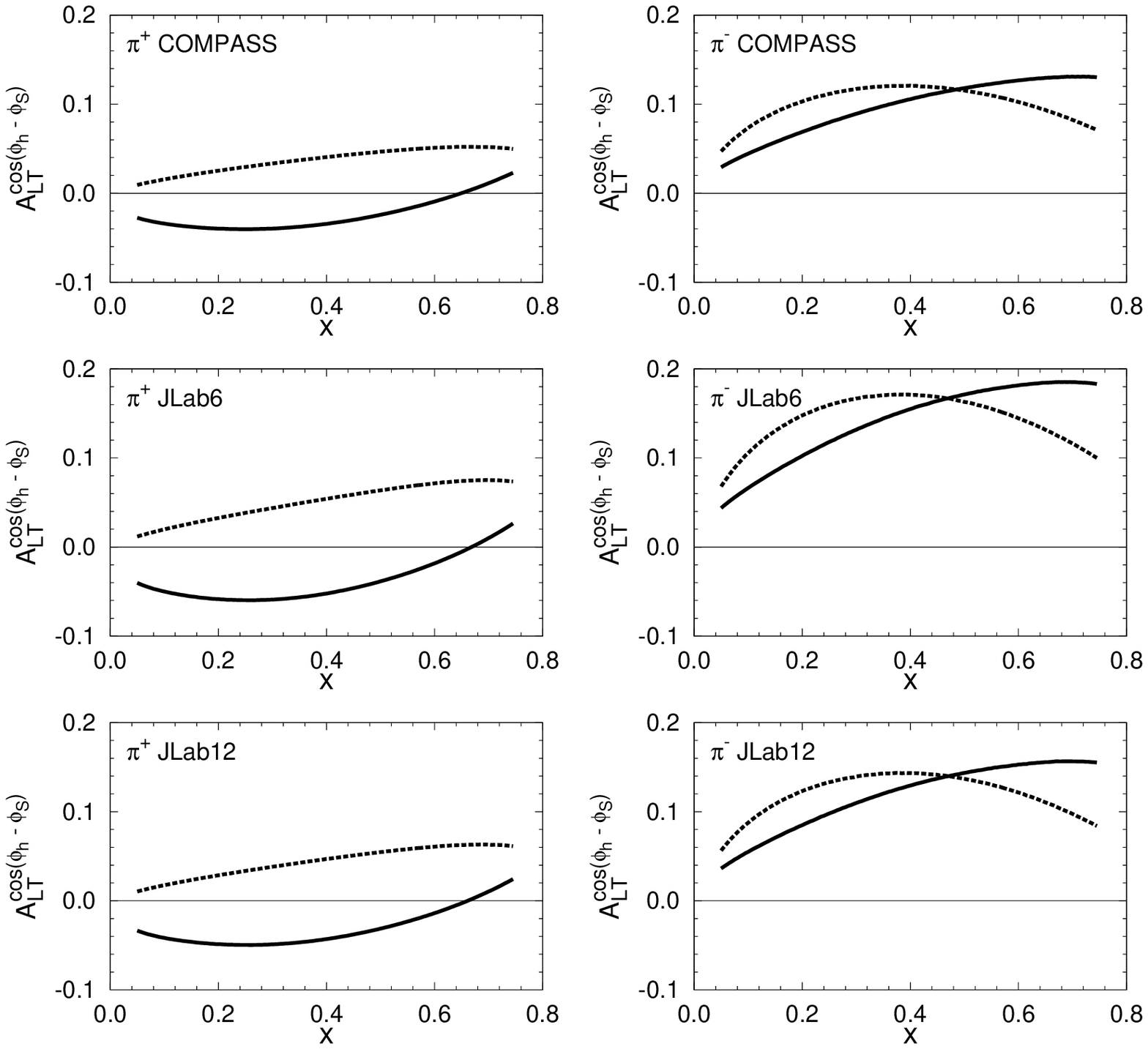}
\caption{\label{fig:g_n}The double spin asymmetry $A_{LT}^{\cos(\phi_h-\phi_S)}$ as a function of $x$ at different kinematics with $Q^2 = 3.0 ~ \mathrm{GeV}^2$ for the neutron target. Dashed curves correspond to approach 1, while solid curves correspond to approach 2.}
\end{figure}

\begin{figure}
\centering
\includegraphics[width=0.80\textwidth]{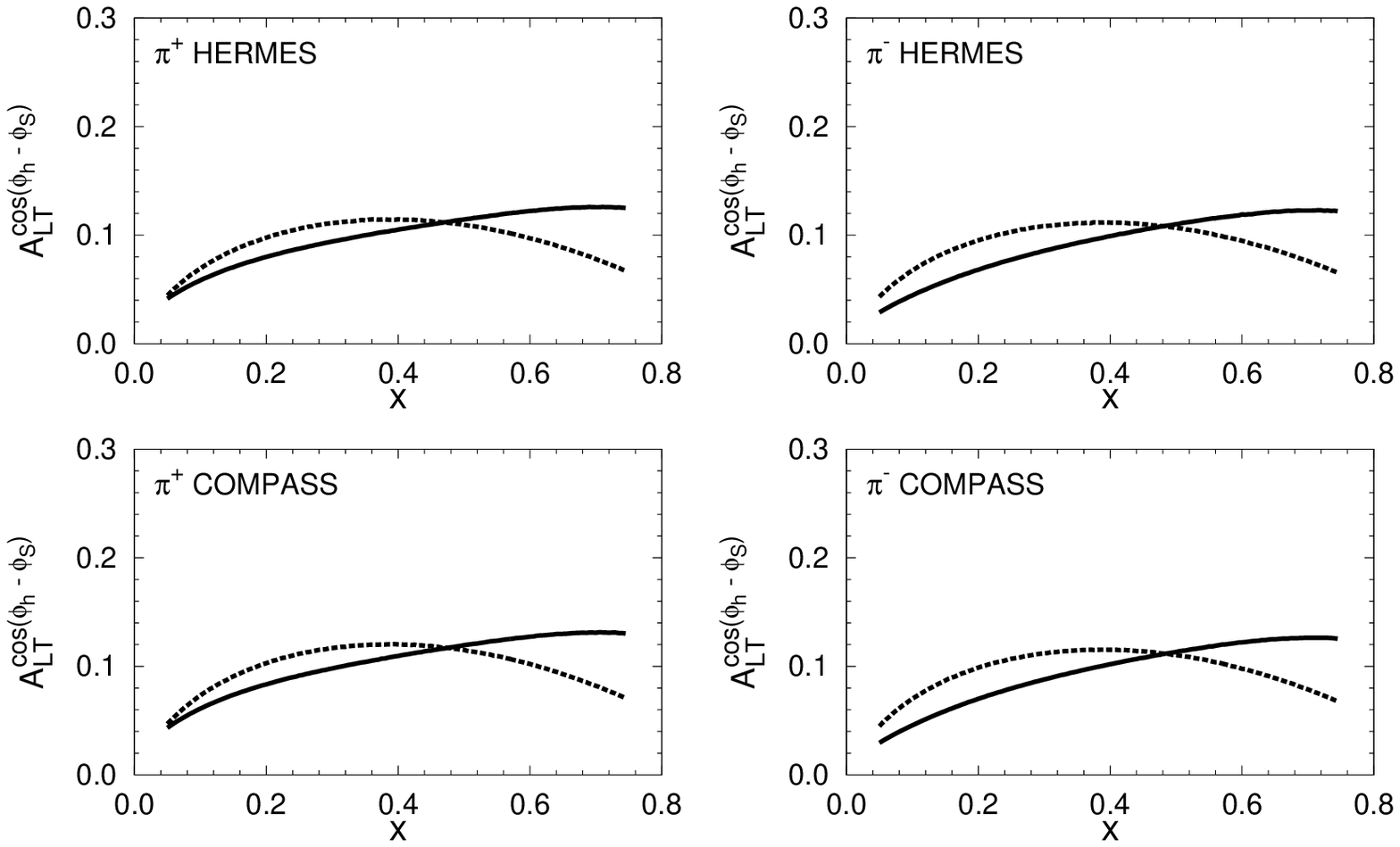}
\caption{\label{fig:g_d}The double spin asymmetry $A_{LT}^{\cos(\phi_h-\phi_S)}$ as a function of $x$ at different kinematics with $Q^2 = 3.0 ~ \mathrm{GeV}^2$ for the deuteron target. Dashed curves correspond to approach 1, while solid curves correspond to approach 2.}
\end{figure}

\begin{figure}
\centering
\includegraphics[width=0.80\textwidth]{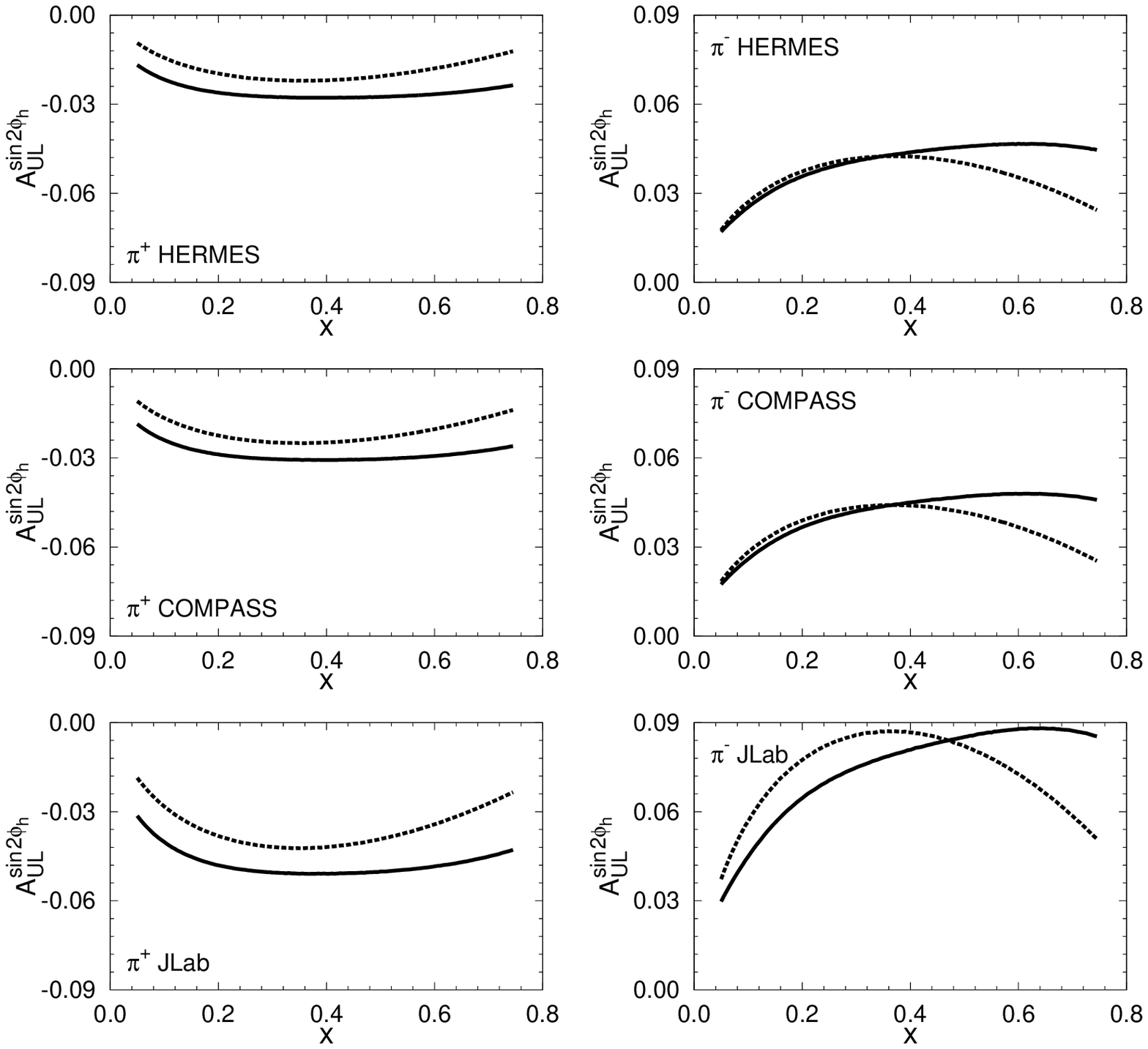}
\caption{\label{fig:h_p}The single spin asymmetry $A_{UL}^{\sin 2\phi_h}$ as a function of $x$ at different kinematics with $Q^2 = 3.0 ~ \mathrm{GeV}^2$ for the proton target. Dashed curves correspond to approach 1, while solid curves correspond to approach 2.}
\end{figure}

\begin{figure}
\centering
\includegraphics[width=0.80\textwidth]{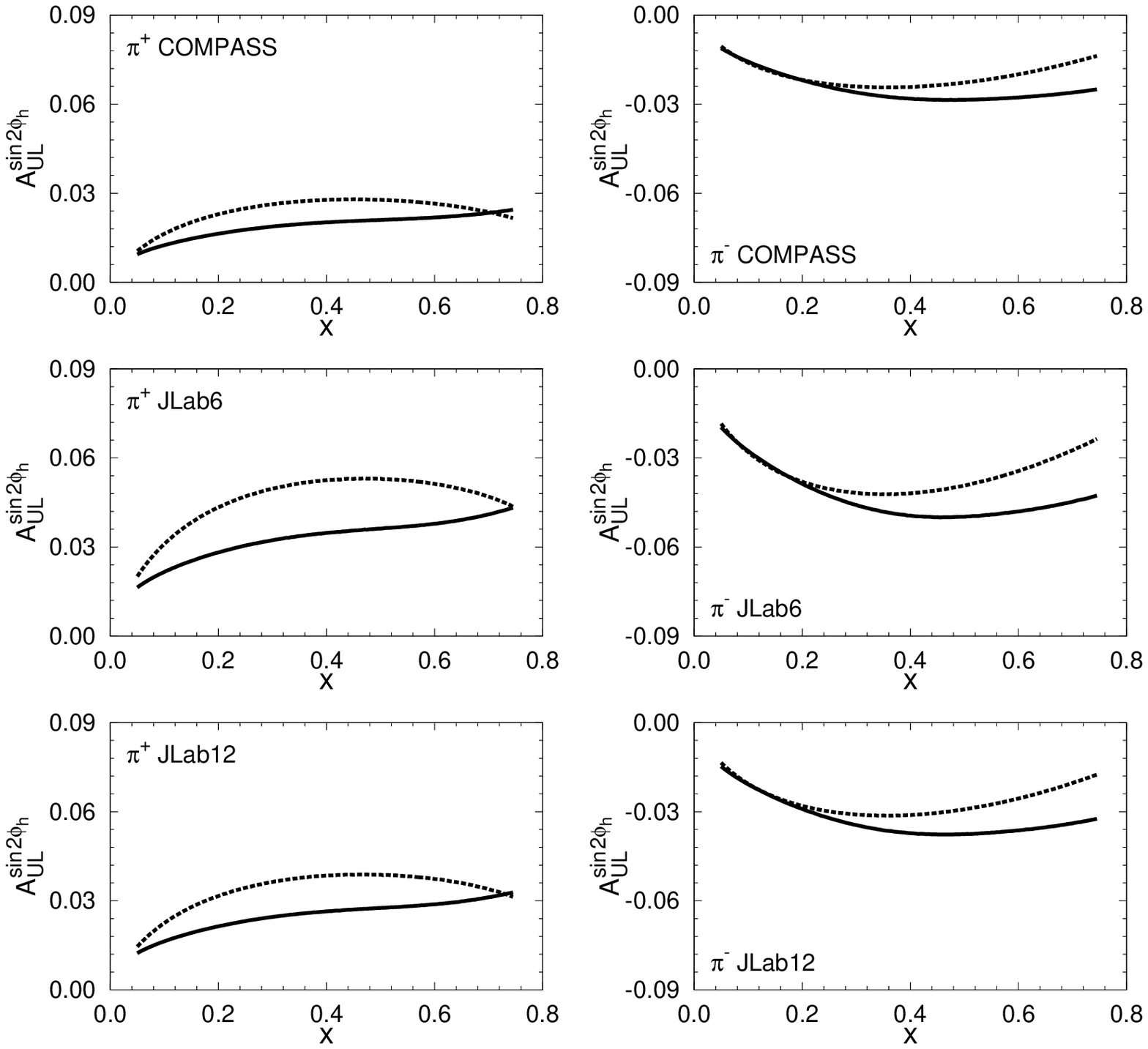}
\caption{\label{fig:h_n}The single spin asymmetry $A_{UL}^{\sin 2\phi_h}$ as a function of $x$ at different kinematics with $Q^2 = 3.0 ~ \mathrm{GeV}^2$ for the neutron target. Dashed curves correspond to approach 1, while solid curves correspond to approach 2.}
\end{figure}

\begin{figure}
\centering
\includegraphics[width=0.80\textwidth]{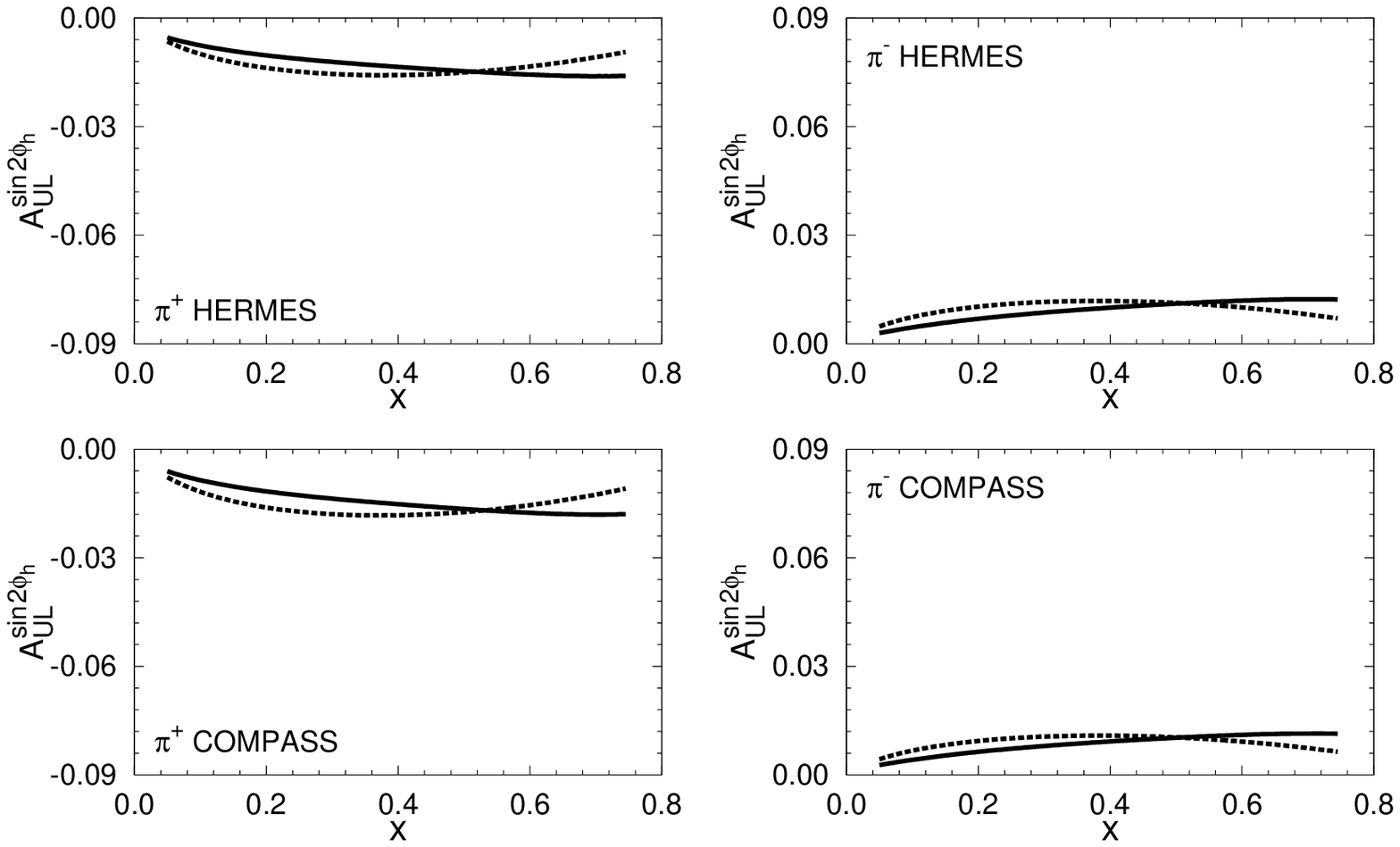}
\caption{\label{fig:h_d}The single spin asymmetry $A_{UL}^{\sin 2\phi_h}$ as a function of $x$ at different kinematics with $Q^2 = 3.0 ~ \mathrm{GeV}^2$ for the deuteron target. Dashed curves correspond to approach 1, while solid curves correspond to approach 2.}
\end{figure}

Now, we present the predictions of the double spin asymmetry
$A_{LT}^{\cos(\phi_h-\phi_S)}$ and the single spin asymmetry
$A_{UL}^{\sin 2\phi_h}$ in SIDIS at different kinematics as shown
in Table.~\ref{table:kin}. Both $\pi^+$ and $\pi^-$ productions of
the proton target in HERMES~\cite{hermes2005,hermes2010},
COMPASS~\cite{compass2005,compass2010} and JLab~\cite{jlab,Gao2010}
experiments, the neutron target in COMPASS and JLab
experiments, and the deuteron target in HERMES and COMPASS
experiments are calculated. The results for the double spin
asymmetry $A_{LT}^{\cos(\phi_h-\phi_S)}$ are shown in
Figs.~\ref{fig:g_p}~\ref{fig:g_n}~\ref{fig:g_d}, and those for
the single spin asymmetry $A_{UL}^{\sin 2\phi_h}$ are shown in
Figs.~\ref{fig:h_p}~\ref{fig:h_n}~\ref{fig:h_d}, respectively.

\begin{table}
\caption{Kinematics at HERMES, COMPASS and JLab} \label{table:kin}
\centering
\scriptsize
\begin{tabular}{c|c|c|c|c|c|c}
\hline\hline
&HERMES&COMPASS&\multicolumn{2}{|c|}{JLab6}&\multicolumn{2}{|c}{JLab12}\\
\cline{4-7}
&&&proton&neutron&proton&neutron\\
\hline
$p_\textrm{lab}$/GeV&27.6&160&6&6&12&12\\
$Q^2$/GeV$^2$&$>1$&$>1$&$>1$&$1.3$--$3.1$&$>1$&$>1$\\
$W^2$/GeV$^2$&$>10$&$>25$&$>4$&$5.4$--$9.3$&$>4$&$>2.3$\\
$x$&$0.023$--$0.4$&$ $&$0.1$--$0.6$&$0.13$--$0.4$&$0.05$--$0.7$&$0.05$--$0.55$\\
$y$&$0.1$--$0.85$&$0.1$--$0.9$&$0.4$--$0.85$&$0.68$--$0.86$&$0.2$--$0.85$&$0.34$--$0.9$\\
$z$&$0.2$--$0.7$&$0.2$--$1$&$0.4$--$0.7$&$0.46$--$0.59$&$0.4$--$0.7$&$0.3$--$0.7$\\
\hline\hline
\end{tabular}
\end{table}

The magnitudes of these asymmetries are significantly larger
compared with those of the single spin asymmetry
$A_{UT}^{\sin(3\phi_h - \phi_S)}$ that we obtained in
Ref.~\cite{She2009}, so a transverse momentum cut is not necessary.
The magnitudes of our results are comparable with those in
Ref.~\cite{Kotzinian2006} but larger than those in
Ref.~\cite{Boffi2009}.

\section{Summary}
We have calculated two of the eight leading twist TMDs, $g_{1T}$ and $h_{1L}^\perp$,
in the light-cone quark-diquark model, and found some interesting
relations among them. They can be measured through SIDIS. The predictions
of the double and single asymmetries in SIDIS related to them have
been presented at HERMES, COMPASS and JLab kinematics for the proton,
neutron and deuteron targets. We expect future experiments can prompt
our understanding of the nucleon spin structure.

\section*{Acknowledgments}
This work is supported by National Natural Science Foundation of
China (Nos. 10721063, 10975003, and 11035003).








\end{document}